\documentstyle[aps,prl,epsfig,twocolumn,times]{revtex}
\draft

\begin{document}
\markright{Garay, Anglin, Cirac, and Zoller.\hspace{3mm}
Sonic analog of gravitational black holes in Bose-Einstein
condensates}
\title{Sonic analog of gravitational black
holes in Bose-Einstein condensates}
\author{L.J.~Garay$^{1,2}$, J.R.~Anglin$^{1,3}$, J.I.~Cirac$^1$,
and P.~Zoller$^{1}$}
\address{$^1$ Institut f\"ur Theoretische Physik, Universit\"at
Innsbruck, Technikerstrasse 25, A-6020 Innsbruck, Austria}
\address{$^2$ Instituto de Matem\'{a}ticas y F\'{\i}sica Fundamental,
CSIC, C/ Serrano
121, E-28006 Madrid, Spain}
\address{$^3$ Institute for Theoretical Atomic and Molecular
Physics, Harvard-Smithsonian Center for Astrophysics,
60 Garden St.,Cambridge
MA 02135}
\date{3 February 2000}

\wideabs{
\maketitle

\begin{abstract}

\indent
 It is shown that,  in dilute-gas
Bose-Einstein condensates,
there exist both dynamically stable and unstable configurations
which, in the hydrodynamic limit, exhibit a behavior
resembling
 that of gravitational black holes. The dynamical
instabilities involve creation of quasiparticle pairs in
positive and negative energy states,  as in the well-known
suggested mechanism for black hole evaporation.
We propose a scheme to generate a  stable sonic black hole in a
ring trap.
\end{abstract}
\pacs{03.75.Fi, 04.70.Dy, 04.80.-y\hfill {\it gr-qc/0002015; Phys.
Rev. Lett. {\bfseries 85}, 4643 (2000)} }}

Many investigations of dilute gas Bose-Einstein condensates are
directed towards experimentally creating nontrivial
configurations of the semiclassical mean field, or to predicting
the properties of such configurations in the presence of quantum
fluctuations. Such problems are hardly peculiar to condensates,
but ultracold dilute gases are so easy to manipulate and
control, both experimentally \cite{BECexp95} and theoretically
\cite{dalfovo+99}, that they may
 allow us to analyze less amenable systems
by analogy. As an essay in such an application of condensates,
in this paper we discuss the theoretical framework and propose
an experiment to create the analog of a black hole in the
laboratory and simulate its radiative instabilities.

The hydrodynamic analog of an event horizon \cite{misner+73} was
suggested originally by Unruh \cite{unruh81} as a more
accessible phenomenon which might shed some light on the Hawking
effect \cite{hawking74} (thermal radiation from black holes,
stationary insofar as backreaction is negligible) and, in
particular, on the role of ultrahigh frequencies
\cite{jacobson91,unruh95,corley+99}. An event horizon for sound
waves appears in principle wherever there is a closed surface
through which a fluid flows inwards at the speed of sound, the
flow being subsonic on one side of the surface and supersonic on
the other. There is a close analogy between sound propagation on
a background hydrodynamic flow, and field propagation in a
curved spacetime; and although hydrodynamics is only a
long-wavelength effective theory for physical (super)fluids, so
also field theory in curved spacetime is to be considered a
long-wavelength approximation to quantum gravity
\cite{unruh95,visser98}. Determining whether and how sonic black
holes radiate sound, in a full calculation beyond the
hydrodynamic approximation or in an actual experiment, can thus
offer some suggestions about black hole radiance and its
sensitivity to high frequency physics.

The basic challenge of our proposal is to keep the trapped
Bose-Einstein gas sufficiently cold and well isolated to
maintain a locally supersonic flow long enough to observe its
intrinsic dynamics. Detecting thermal phonons radiating from the
horizons would obviously be a difficult additional problem,
since such radiation would be indistinguishable from many other
possible heating effects. This further difficulty does not arise
in our proposal, however, because the black-hole radiation we
predict is, unlike Hawking radiation, not quasistationary, but
grows exponentially under appropriate conditions. It should
therefore be observable in the next generation of atom traps.

A Bose-Einstein condensate is the ground state of a second
quantized many body Hamiltonian for $N$ interacting bosons
trapped by an external potential $V_{\rm ext}({\mathbf{x}})$
\cite{dalfovo+99}. At zero temperature, when the number of atoms
is large and the atomic interactions are sufficiently small,
almost all the atoms are in the same single-particle quantum
state $\Psi({\mathbf{x}},t)$, even if the system is slightly
perturbed. The evolution of $\Psi$ is then given by the
well-known Gross-Pitaevskii equation
$$
i\hbar\partial_t\Psi=\left(-\frac{\hbar^2}{2m}\nabla^2+ V_{\rm
ext}+\frac{4\pi a\hbar^2}{m}|\Psi|^2\right)\Psi ,
$$
where $m$ is the mass of the atoms,  $a$ is the
scattering length, and we normalize
to the total number of atoms $\int d^3{\mathbf{x}}
|\Psi({\mathbf{x}},t)|^2=N$.

Our purposes do not require solving the Gross-Pitaevskii
equation with some given external potential $V_{\rm
ext}({\mathbf{x}})$; our concern is the propagation of small
collective perturbations of the condensate, around a background
stationary state
$
\Psi_s({\mathbf{x}},t)=\sqrt{\rho({\mathbf{x}})}
e^{i\vartheta({\mathbf{x}})} e^{-i\mu t/\hbar},
$
where $\mu$ is the chemical potential. Thus it is only necessary
that it be possible, in any external potential that can be
generated, to create a condensate in this state. Many realistic
techniques for ``quantum state engineering,'' to create designer
potentials and bring condensates into specific states, have been
proposed, and even implemented successfully \cite{cornell99};
our simulations indicate that currently known techniques should
suffice to generate the condensate states that we propose.

Perturbations about the stationary state
$\Psi_s({\mathbf{x}},t)$ obey the Bogoliubov system of two
coupled second order differential equations. Within the regime
of validity of the hydrodynamic (Thomas-Fermi) approximation
\cite{dalfovo+99}, these two equations for the density
perturbation $\varrho$ and the phase perturbation $\phi$ in
terms of the local speed of sound
$
c({\mathbf{x}})\equiv\frac{\hbar}{m}\sqrt{4\pi
a\rho({\mathbf{x}})},
$
and the background stationary velocity
$
{\mathbf{v}}\equiv\frac{\hbar}{m}\nabla\vartheta
$
read
$$
\dot\varrho=-\nabla(\frac{m}{4\pi a\hbar}c^2\nabla\phi+{\mathbf{v}}\varrho)
,\quad
\dot\phi=-{\mathbf{v}}\nabla\phi -\frac{4\pi a\hbar}{m}\varrho.
$$
Furthermore, low frequency perturbations are essentially just
waves of (zero) sound. Indeed, the Bogoliubov equations
 may be reduced to a single second order equation for the
condensate phase perturbation $\phi$. This differential equation
has the form of a relativistic wave equation
$\partial_\mu(\sqrt{-g}g^{\mu\nu}\partial_\nu\phi)=0$, with
$g=\det g_{\mu\nu}$, in an effective curved spacetime with the
metric $g_{\mu\nu}$ being entirely determined by the local speed
of sound $c$ and the background stationary velocity
${\mathbf{v}}$. Up to a conformal factor, this effective metric
has the form
$$
(g_{\mu\nu})=\left(\begin{array}{cc} -(c^2- {\mathbf{v}}^2) &
-{\mathbf{v}}^{\rm T}\\ -{\mathbf{v}} & {\mathbf{1}}
\end{array}\right).
$$

This class of metrics can possess event horizons. For instance,
if an effective sink for atoms is generated at the center of a
spherical trap (such as by an atom laser out-coupling technique
\cite{andrews97}), and if the radial potential profile is
suitably arranged, we can produce densities $\rho(r)$ and flow
velocities ${\mathbf{v}}({\mathbf{x}})=-v(r){\mathbf{r}}/r$ such
that the quantity $c^2-{\mathbf{v}}^2$ vanishes at a radius
$r=r_h$, being negative inside and positive outside. The sphere
at radius $r_h$ is a sonic event horizon completely analogous to
those appearing in gravitational black holes, in the
sense that sonic perturbations cannot propagate through this
surface in the outward direction
\cite{unruh81,unruh95,visser98}. The physical mechanism of the sonic
black hole is quite simple: inside the horizon, the background
flow speed $v$ is larger than the local speed of sound $c$, and
so sound waves are  dragged inwards.

In fact there are two conditions which must hold for this
dragged sound picture to be accurate. Wavelengths larger than
the black hole itself will of course not be dragged in, but
merely diffracted around it. And perturbations must have
wavelengths $\lambda\gg 2\pi\xi,\ 2\pi\xi/\sqrt{|1-v/c|}$, where
$\xi({\mathbf{x}})\equiv\hbar/[mc({\mathbf{x}})]$ is the local
healing length.
Otherwise they do not behave as sound waves since they lie
outside the regime of validity of the hydrodynamic
approximation. These short-wavelength modes must be described by
the full Bogoliubov equations, which allow signals to propagate
faster than the local sound speed, and thus permit escape from
sonic black holes.  Even if such an intermediate
range of wavelengths does exist, the modes outside it may still affect the
stability of the black hole as discussed below.

As it stands, this description is incomplete. The condensate
flows continually inwards and therefore at $r=0$ there must be a
sink that takes atoms out of the condensate. Otherwise, the
continuity equation $\nabla (\rho{\mathbf{v}})=0$, which must
hold for stationary configurations, will be violated. We have
analyzed several specific systems which may be suitable
theoretical models for future experiments, and have found that
the qualitative behavior is analogous in all of them. Black
holes which require atom sinks are both theoretically and
experimentally more involved, however; moreover, maintaining a
steady transonic flow into a sink may require either a very
large condensate or some means of replenishment. We will
therefore discuss here an alternative configuration which may be
experimentally more accessible and whose description is
particularly simple: a condensate in a very thin ring that
effectively behaves as a periodic one-dimensional system. Under
conditions that we will discuss, the supersonic region in a ring
may be bounded by two horizons: a black hole horizon through
which phonons cannot exit, and a ``white hole'' horizon through
which they cannot enter.

In a sufficiently tight ring-shaped external potential of radius $R$, motion
in radial ($r$) and axial ($z$) cylindrical coordinates is
effectively frozen. We can then write the wave function as
$
\Psi(z,r,\theta,\tau)=f(z,r)\Phi(\theta,\tau)
$
and normalize $\Phi$ to the number of atoms in the condensate
$\int_0^{2\pi}d\theta|\Phi(\theta)|^2= N$, where with the azimuthal
coordinate $\theta$ we have introduced the  dimensionless time
$\tau=\frac{\hbar}{mR^2}t$.
The Gross-Pitaevskii equation thus becomes effectively
one-dimensional:
\begin{equation}
i\partial_\tau\Phi=\left(-\frac{1}{2}\partial_\theta^2+{\cal
V}_{\rm ext}+\frac{{\cal U}}{N} |\Phi|^2\right)\Phi,
\label{eq:gp-ring}
\end{equation}
where
$
{\cal U} \equiv 4\pi a N R^2 \int dzdrr |f(z,r)|^4
$
and ${\cal V}_{\rm ext}(\theta)$ is the dimensionless effective
potential (in which we have already included the chemical
potential) that results from the dimensional reduction. The
stationary solution can then be written as
$
\Phi_s(\theta,\tau)=\sqrt{\rho(\theta)} e^{i\int\!d\theta
v(\theta)}
$
and the local dimensionless angular speed of sound as
$c(\theta)=\sqrt{{\cal U}\rho(\theta)/N}$. Periodic boundary
conditions around the ring require the ``winding number''
$
w\equiv\frac{1}{2\pi}\int_0^{2\pi}d\theta v(\theta)
$
to be an integer.

The qualitative behavior of horizons in a ring
is well represented by the two-parameter family of condensate
densities
$$
\rho(\theta)=\frac{N}{2\pi}(1+b\cos\theta),
$$
where $b\in[0,1]$. Continuity, $\partial_\theta(\rho v)=0$, then
determines the dimensionless flow-velocity field
$$
v(\theta)=\frac{{\cal U}w\sqrt{1-b^2}}{2\pi c(\theta)^2},
$$
which depends on $w$ as a third discrete independent parameter.
Requiring that $\Phi_s(\theta,\tau)$ be a stationary solution to
Gross-Pitaevskii equation then determines how the trapping
potential must be modulated as a function of $\theta$. All the
properties of the condensate, including whether and where it has
sonic horizons, and whether or not they are stable, are thus
functions of ${\cal U}$, $b$ and $w$. For instance, if we
require that the horizons be located at $\theta_h=\pm\pi/2$,
which imposes the relation ${\cal U}=2\pi w^2(1-b^2)$, then we
must have $c^2-v^2$ positive for $\theta\in (-\pi/2,\pi/2)$,
zero at $\theta_h=\pm\pi/2$, and negative otherwise, provided
that ${\cal U}<2\pi w^2$. The further requirement that
perturbations on wavelengths shorter than the inner and the
outer regions are indeed phononic implies ${\cal U}\gg2\pi$,
which in turn requires $w\gg1$ and $1\gg b\gg 1/w^2$. In fact,
detailed analysis shows that $w\gtrsim 5$ is sufficient.

A black hole solution should also be stable over sufficiently
long time scales in order to be physically realizable. Since
stability must be checked for perturbations on all wavelengths,
the full Bogoliubov \cite{dalfovo+99} spectrum must be
determined. For large black holes within infinite condensates,
this Bogoliubov problem may be solved using WKB methods that
closely resemble those used for solving relativistic field
theories in true black hole spacetimes \cite{corley+99}. The
results are also qualitatively similar to those we have found
for black holes in finite traps, where we have resorted to
numerical methods because, in these cases, WKB techniques may
fail for just those modes which threaten to be unstable.

Our numerical approach for our three-parameter family of
black/white holes in the ring-shaped condensate has been to
write the Bogoliubov equations in discrete Fourier space, and
then truncate the resulting infinite-dimensional eigenvalue
problem. Writing the wave funtion as $\Phi=\Phi_s
+\varphi e^{i\int d\theta v(\theta)}$,
decomposing the perturbation $\varphi$ in discrete modes
\begin{eqnarray}
\varphi(\theta,\tau)=&&\sum_{\omega,n} e^{-i\omega \tau}
e^{in\theta}
A_{\omega,n}u_{\omega,n}(\theta)
\nonumber\\
&&{}+ e^{i\omega^* \tau} e^{-in\theta}
A_{\omega,n}^*v_{\omega,n}^*(\theta),
\nonumber
\end{eqnarray}
 and substituting into the Gross-Pitaevskii equation, we obtain
the following equation for the modes $u_{\omega,n}$ and
$v_{\omega,n}$:
$$
\omega \left(\begin{array}{c} u_{\omega,n} \\ v_{\omega,n}
\end{array}\right)=
\sum_p \left(\begin{array}{cc}
h^+_{np} & f_{np}\\ -f_{np} & h^-_{np}
\end{array}\right)
\left(\begin{array}{c} u_{\omega,p} \\ v_{\omega,p}
\end{array}\right).
$$
In this equation,
\begin{eqnarray}
f_{np}&=&\frac{{\cal U}}{2\pi}\left(\delta_{n,p}+ \frac{b}{2}
\delta_{n,p+1} + \frac{b}{2}
\delta_{n,p-1}\right),\nonumber\\
h^\pm_{np}&=&\frac{1}{2}(n+p)w\sqrt{1-b^2}\alpha_{n-p}
\nonumber\\
&\pm&\left(f_{np}+\frac{4n^2-1}{8}\delta_{n,p}+
\frac{1-b^2}{8}\beta_{n-p}
\right)
,
\nonumber
\\ 
\alpha_i&=&\sum_{j\geq |i|,\ i+j\ {\rm even}}^\infty
\left(\frac{-b}{2}\right)^j\left(\begin{array}{c} j\\
(i+j)/{2}\end{array}\right),\nonumber\\
\beta_i&=&\sum_{j\geq |i|,\ i+j\ {\rm even}}^\infty
\left(\frac{-b}{2}\right)^j\left(\begin{array}{c} j\\
(i+j)/{2} \end{array}\right) (j+1).\nonumber
\end{eqnarray}
 Eliminating Fourier components above a sufficiently high
cutoff $Q$ has negligible effect on possible instabilities,
which can be shown to occur at relatively long wavelengths. The
numerical solution to this eigenvalue equation, together with
the normalization condition $\int d\theta
(u^*_{\omega^*,n}u_{\omega',n'}- v^*_{\omega^*,n}v_{\omega',n'})
=\delta_{nn'}\delta_{\omega\omega'}$, provides the allowed
frequencies. Real negative eigenfrequencies for modes of
positive norm are always present, which means that black hole
configurations are energetically unstable, as expected. This
feature is inherent in supersonic flow, since the speed of sound
is also the Landau critical velocity. In a sufficiently cold and
dilute condensate, however, the time scale for dissipation may
in principle be made very long, and so these energetic
instabilities need not be problematic \cite{shlyapnikov}.

\begin{figure}
\begin{center}
\epsfig{file=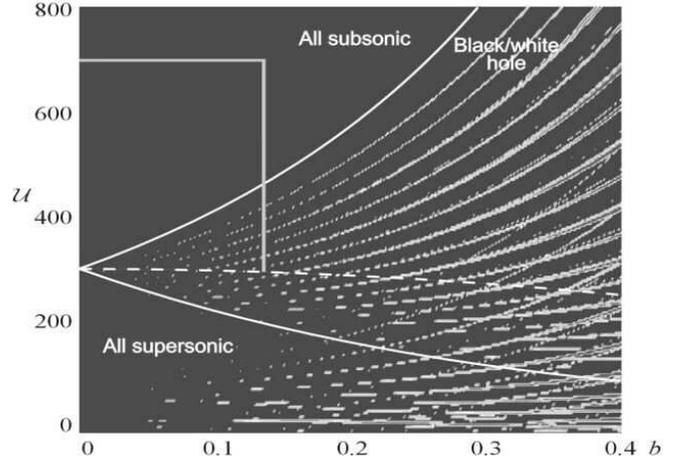,width=\columnwidth}
\end{center}
\caption{Stability diagram for winding number $w=7$. Solid dark-grey
areas represent the regions of stability. Smaller plots at
higher resolution confirm that the unstable ``fingers'' are
actually smooth and unbroken. Points on the dashed curve are
states with horizons at $\pm\pi/2$, so that the black/white hole
fills half the ring.}
\label{fig:stab}
\end{figure}

More serious are dynamical instabilities, which occur for modes
with complex eigenfrequencies and are genuine physical
phenomena. For sufficently high values of the cutoff (e.g.,
$Q\geq 25$ in our calculations), the complex eigenfrequencies
obtained from the truncated eigenvalue problem become
independent of the cutoff within the numerical error. The
existence and rapidity of dynamical instabilities depend
sensitively on $({\cal U},b,w)$. For instance, see
Fig.~\ref{fig:stab} for a contour plot of the maximum of the
absolute values of the imaginary parts of all eigenfrequencies
for $w=7$, showing that the regions of instability are long,
thin fingers in the $({\cal U},b)$ plane. Not shown in the
figure is the important fact that the size of the imaginary
parts, which gives the rate of the instabilities, increases
starting from zero, quite rapidly with $b$, although they remain
small as compared with the real parts.

The stability diagram of Fig.~\ref{fig:stab} suggests a strategy
for creating a sonic black hole from an initial stable state.
Within the upper subsonic region, the vertical axis $b=0$
corresponds to a homogeneous persistent current in a ring, which
can in principle be created using different techniques
\cite{dum98}. Gradually changing ${\cal U}$ and $b$, it is
possible to move from such an initial state to a black/white
hole state, along a path lying almost entirely within the stable
region, and passing only briefly through instabilities where
they are sufficiently small to cause no difficulty.

Indeed, we have simulated this process of adiabatic creation of
a sonic black/white hole by solving numerically
(using the split operator method)
the time-dependent Gross-Pitaevskii equation~(\ref{eq:gp-ring})
that provides the
evolution of the condensate when the parameters of the trapping
potential change so as to move the condensate state along
various paths in parameter space. One of these paths is shown in
Fig.~\ref{fig:stab} (light-grey solid line): we start with a current at
$w=7$, $b=0$, and sufficiently high ${\cal U}$; we then increase
$b$ adiabatically keeping ${\cal U}$ fixed until an appropriate
value is reached; finally, keeping $b$ constant, we decrease
${\cal U}$ adiabatically (which can be physically implemented by
decreasing the radius of the ring trap), until we meet the
dashed contour for black holes of comfortable size. Our
simulations confirm that the small instabilities which briefly
appear in the process of creation do not disrupt the adiabatic
evolution. The final quantum state of the condensate, obtained
by this procedure, indeed represents a stable black/white hole.
We have further checked the stability of this final
configuration by numerically solving the Gross-Pitaevskii
equation~(\ref{eq:gp-ring}) for very long periods of time
(as compared with any
characteristic time scale of the condensate) and for fixed
values of the trap parameters. This evolution reflects the fact
that no complex frequencies are present, as predicted from the
mode analysis, and that the final state is stationary.

Once the black/white hole has been created, one could further
change the parameters $({\cal U},b)$ so as to move between the
unstable ``fingers'' into a stable region of higher $b$ (a
deeper hole); or one could deliberately enter an unstable
region. In the latter case, the black hole should disappear in
an explosion of phonons, which may be easy to detect
experimentally. Such an event might be related to the
evaporation process suggested for real black holes in the sense
that pairs of quasiparticles are created near the horizon in
both positive and negative energy modes. The Hermiticity of the
Bogoliubov Hamiltonian implies that eigenmodes with complex
frequencies appear always in dual pairs, whose frequencies are
complex conjugate. In the language of second quantization, the
linearized Hamiltonian for each such pair has the form
$$
H=\sum_{n}( \omega A^\dag_{\omega^*,n} A_{\omega,n}
+\omega^* A^\dag_{\omega,n} A_{\omega^*,n}),
$$
and the only nonvanishing commutators among these operators are
$[A_{\omega,n},A^\dag_{\omega^*,n'}]=\delta_{nn'}$. It is then
clear that none of these operators is actually a harmonic
oscillator creation or annihilation operator in the usual sense.
However, the linear combinations (note that
$A^\dag_{\omega^*,n}\neq A^\dag_{\omega,n}$)
$$
a_{n}=\frac{1}{\sqrt 2}(A_{\omega,n}+A_{\omega^*,n})\ ,\qquad
b_{n}=\frac{i}{\sqrt 2}(A^\dag_{\omega,n}+A^\dag_{\omega^*,n})
$$
and their Hermitian conjugates are true annihilation and
creation operators, with the standard commutation relations, and
in terms of these the Bogoliubov Hamiltonian becomes
$$
H = \sum_n\left[{\rm Re}(\omega)(a^\dag_{n} a_{n} -
b^\dag_{n} b_{n}) - {\rm
Im}(\omega)(a^\dag_{n} b^\dag_{n}+
a_{n} b_{n})\right]\;,
$$
which obviously leads to self-amplifying creation of positive and negative
frequency pairs.
Evaporation
through an exponentially self-amplifying instability is not
equivalent, however, to the usual kind of Hawking radiation
\cite{corley+99}; this issue will be discussed in detail
elsewhere.

Trapped bosons at ultralow temperature can provide an analog to
a black-hole spacetime. Similar analogs have been proposed in
other contexts, such as superfluid helium \cite{ruutu+96}, solid
state physics \cite{reznik97}, and optics \cite{leonhardt+00};
but the outstanding recent experimental progress in cooling,
manipulating and controlling atoms \cite{cornell99} makes
Bose-Einstein condensates an especially powerful tool for this
kind of investigation. We have analyzed in detail the case of a
condensate in a ring trap, and proposed a realistic scheme for
adiabatically creating stable sonic black/white holes.

We thank the Austrian Science Foundation and the European Union
TMR networks ERBFMRX--CT96--0002 and ERB--FMRX--CT96--0087.

{\it Note added.---} Further details as well as the study of
cigar shaped condensates with atom sinks at the center will
appear in Ref. \cite{bhbec-pra}.


\begin{references}
\bibitem{BECexp95} M. H. Anderson {\it et al.}, {Science} {\bfseries
269}, 198 (1995); K. B. Davis et al., {Phys. Rev. Lett.}
{\bfseries 75}, 3969 (1995).

\bibitem{dalfovo+99}
See, e.g., F. Dalfovo {\it et al.},  {Rev. Mod. Phys.} {\bfseries 71}, 463
(1999).

\bibitem{misner+73} C. W. Misner, K. S. Thorne, and J. A. Wheeler,
{\itshape Gravitation} (Freeman, San Francisco, 1973).

\bibitem{unruh81}
W.G. Unruh, {Phys. Rev. Lett.} {\bfseries 46}, 1351
(1981).

\bibitem{hawking74} S. W. Hawking, {Nature} {\bfseries 248},
30 (1974); {Commun. Math. Phys.} {\bfseries 43}, 199
(1975).


\bibitem{jacobson91}
T. Jacobson, {Phys. Rev.} D {\bfseries 44}, 1731 (1991).

\bibitem{unruh95}
W. G. Unruh, {Phys. Rev.} D {\bfseries 51}, 2827 (1995).

\bibitem{corley+99}
S. Corley and T. Jacobson, {Phys. Rev.} D {\bfseries 59}, 4011
(1999); S. Corley, {Phys. Rev.} D {\bfseries 57}, 6280 (1998).

\bibitem{visser98}
M. Visser, {Phys. Rev. Lett.} {\bfseries 80}, 3436
(1998); {Class. Quant. Grav.} {\bfseries 15}, 1767
(1998).



\bibitem{cornell99}
M. R. Matthews {\it et al.}, {Phys.
  Rev. Lett.} {\bfseries  83},  2498  (1999);
L. Denget {\it et al.}, {Nature} {\bfseries 398}, 218 (1999); S.
Burger {\it et al.}, {Phys. Rev. Lett.} {\bfseries 83}, 5198
(1999).

\bibitem{andrews97}
M. R. Andrews {\it et al.}, {Science} {\bfseries 275}, 637
(1997); I. Bloch, T. W. H\"ansch, and T. Esslinger, {Phys. Rev.
Lett.} {\bfseries 82}, 3008 (1999); E. W. Hagley
 {\it et al.}, {Science} {\bfseries 283}, 1706 (1999).

\bibitem{shlyapnikov}
P. O. Fedichev and G. V. Shlyapnikov, {Phys. Rev.} A
{\bfseries 60}, R1779 (1999).

\bibitem{dum98} R. Dum {\it et al.},
{Phys. Rev. Lett.} {\bfseries 80}, 2972 (1998); J. Williams and
M. Holland, {Nature} {\bfseries 401}, 568 (1999).

\bibitem{ruutu+96}
V. M. H. Ruutu {\it et al.}, {Nature} {\bfseries 382}, 334
(1996);
 T. A. Jacobson and G. E. Volovik, {Phys. Rev.} D
{\bfseries 58}, 4021 (1998); G. E. Volovik, {Pisma Zh. Eksp.
Teor. Fiz.} {\bfseries 69}, 662 (1999); JETP Lett. {\bfseries
69}, 705 (1999).

\bibitem{reznik97}
B. Reznik, gr-qc/9703076.

\bibitem{leonhardt+00}
U.~Leonhardt and P.~Piwnicki, {Phys. Rev. Lett.}
{\bfseries 84}, 822 (2000).

\bibitem{bhbec-pra} L. J. Garay {\it et al.}, Phys. Rev. A (2000)
to be published.

\end{references}
\end{document}